\documentstyle[12pt,epsfig]{article}

\newcommand {\blackboardrrm}
{\mathchoice{\rm I\kern-0.21 em{R}}
{\rm I\kern-0.21 em{R}}{\rm I\kern-0.19 em{R}}{\rm I\kern-0.19 em{R}}}

\newcommand {\blackboardzrm}
{\mathchoice{\rm Z\kern-0.32 em{Z}}
{\rm Z\kern-0.32 em{Z}}{\rm Z\kern-0.28 em{Z}}{\rm Z\kern-0.28 em{Z}}}

\newcommand {\be}     	{\begin{equation}}
\newcommand {\ee}     	{\end{equation}}
\newcommand {\lan}    	{\langle}
\newcommand {\ran}    	{\rangle}

\renewcommand {\d}    	{{\mbox d}}

\begin{document}

\title{Violation of the Fluctuation Dissipation Theorem in Finite
 Dimensional Spin Glasses}

\author{Enzo Marinari$^{(a)}$, Giorgio Parisi$^{(b)}$, \\[0.5em]
Federico Ricci-Tersenghi$^{(b)}$
and Juan J. Ruiz-Lorenzo$^{(b)}$\\[0.6em]
$^{(a)}$
{\small Dipartimento di Fisica and Infn, Universit\`a di Cagliari}\\
{\small Via Ospedale 72, 07100 Cagliari (Italy)}\\[0.3em]
{\small \tt marinari@ca.infn.it}\\[0.5em]
$^{(b)}$
{\small Dipartimento di Fisica and Infn, Universit\`a di Roma {\em La
Sapienza}}\\
{\small P. A. Moro 2, 00185 Roma (Italy)}\\[0.3em]
{\small \tt giorgio.parisi@roma1.infn.it}\\
{\small \tt riccife@chimera.roma1.infn.it}\\
{\small \tt ruiz@chimera.roma1.infn.it}\\[0.5em]
}

\date{October 10, 1997}

\maketitle

\begin{abstract}

We study the violation of the fluctuation-dissipation theorem in the
three and four dimensional Gaussian Ising spin glasses using on and
off equilibrium simulations.  We have characterized numerically the
function $X(C)$ that determine the violation and we have studied its
scaling properties.  Moreover we have computed the function $x(C)$
which characterize the breaking of the replica symmetry directly from
equilibrium simulations.  The two functions are numerically equal and
in this way we have established that the conjectured connection
between the violation of fluctuation dissipation theorem in the
off-equilibrium dynamics and the replica symmetry breaking at
equilibrium holds for finite dimensional spin glasses. These results
point to a spin glass phase with spontaneously broken replica symmetry
in finite dimensional spin glasses.

\end{abstract}  

\thispagestyle{empty}
\newpage

\section{\protect\label{S_INT}Introduction}

One of the characteristics of the disordered systems at low
temperatures (and also of real glasses) is that its approach to
equilibrium is very slow, and it is difficult to study equilibrium
properties. Obviously in the high temperature regime there is a fast
approach to the equilibrium.

Due to these large time scales, the out of equilibrium regime becomes
very important since in nature the system remains in this regime long
times (minutes, days or even years). From the theoretical point of
view it is interesting to develop a theory to describe this
regime~\cite{OUT}.

In this paper we will only discuss the low temperature phase
(i.e. below the phase transition point of the system) and center the
discussion on Ising spin glasses above their lower critical dimension
(that lies clearly below three dimensions \cite{BOOK}).

In the disordered case and using the Mean Field approximation
(i.e. infinite range interactions) Cugliandolo and Kurchan have
derived a generalization of the fluctuation dissipation theorem (FDT)
that involves a new function (denoted by $X$) that determines
multiplicatively (see below) the off-equilibrium regime. In the
equilibrium regime $X=1$ and we recover FDT. It is possible to link
this $X$ function with the static (equilibrium) function $x(q)$ (or
its inverse $q(x)$) that appears in the replica symmetry breaking
solution of infinite dimensional spin glasses~\cite{MEPAVI}.

Unfortunately a direct check of this relation between static and
dynamic in realistic models (like finite dimensional spin glasses)
still lacks. One of the goals of this paper is to check this
static-dynamic link in finite dimensional spin glasses.

The crucial point of the relation between the static and dynamic is
that it is possible to compute the complete functional form of the
order parameter (the order parameter is a number in ordered system but
it is a function, $q(x)$, in infinite dimensional spin glass) using
off-equilibrium simulations. Violations of the FDT relations have been
reported for fragile glasses \cite{GIORGIO_GLASS}, but in this case
the corresponding equilibrium computations are still missing.

On the other hand, equilibrium simulations of the three dimensional
spin glasses are very hard and difficult~\cite{BOOK}. It is
interesting to examine different methods than can provide us
equilibrium information without to perform (expensive) equilibrium
simulations. These methods exist and are based on off-equilibrium
simulations (see for instance \cite{MAPARURI,4DIM,6DIM}). They have
been used, for example, in the four dimensional Ising spin glass to
extract the Edward-Anderson order parameter~\cite{4DIM}. One clear
advantage is that, after a fast initial transient, no thermalization
is needed.  Another advantage is that it is possible to simulate large
lattices and so the final results have irrelevant finite size effects.

Following this philosophy we have computed the order parameter
function\footnote{We have computed directly an integrated version of
the order parameter $P(q)$, from which $P(q)$ can be reobtained by
double derivative.} both from off-equilibrium numerical simulations
and from equilibrium ones and we have obtained an impressive agreement
between both approaches that confirm the link between the static and
dynamics in finite dimensional spin glass and provide us with
off-equilibrium numerical methods to compute static quantities like
the probability distribution of the overlap ($P(q)=\d x/\d q$) and the
Edward-Anderson order parameter.

We have simulated the Gaussian Ising spin glass in three and four
dimensions on a hypercubic lattice with periodic boundary
conditions. The Hamiltonian of the system is given by

\be
{\cal H}=-\sum_{<ij>} \sigma_i J_{ij} \sigma_j\ .
\ee
By $<ij>$ we denote the sun over nearest neighbor pairs. The
$J_{ij}$ are Gaussian 
variables with zero mean and unit variance.

The plan of the paper is the following. In the next section we fix the
notation and develop some analytical results. In sections three and
four we show the numerical simulation for the three and four
dimensional Ising spin glasses (respectively). Finally we present the
conclusions.

\section{\protect\label{S_MOD}Analytical Results}

Let us fix our notations. We will study the quantity $A(t)$ that
depends on the local variables of our original Hamiltonian ($\cal H$).
We can define the associate auto-correlation function

\be
C(t,t^\prime) \equiv \lan A(t)  A(t^\prime)  \ran \ ,
\label{auto}
\ee
and the response function

\be
R(t,t^\prime) \equiv \left. \frac{\delta A(t)}{\delta
\epsilon(t^\prime)}\right|_{\epsilon=0} \ ,
\label{res}
\ee
where we have assumed that the original Hamiltonian has been perturbed
by a term

\be
{\cal H}^\prime= {\cal H} + \int \epsilon(t) A(t) \ .
\ee
In the dynamical framework assuming time translational invariance it
is possible to derive the fluctuation-dissipation theorem (thereafter
FDT), that reads as

\be
R(t,t^\prime)=\beta \theta(t-t^\prime) \frac{\partial
C(t,t^\prime)}{\partial t^\prime} \ .
\protect\label{FDT}
\ee
As we are interested in spin models we have chosen
$A(t)=\sigma_i(t)$.  The brackets 
$\lan (\cdot\cdot \cdot) \ran $ in eq. (\ref{auto})
imply here a double average, one over the dynamical process and a
second over the disorder.

The fluctuation dissipation theorem holds in the equilibrium regime,
but in the early regimes of the dynamic we expect a breakdown of its
validity. Mean Field studies \cite{CUKU} suggest the following
modification of the FDT:

\begin{equation}
R(t,t^\prime)=\beta X(t,t^\prime) \theta(t-t^\prime) \frac{\partial
C(t,t^\prime)}{\partial t^\prime} \ .
\end{equation}
It has also been suggested in \cite{CUKU,BCKP} that the function
$X(t,t^\prime)$ is a function of the autocorrelation function:
$X(t,t^\prime)=X(C(t,t^\prime))$.  We can then write the following
generalization of FDT, which should hold in early times of the dynamics, 
the off-equilibrium
fluctuation-dissipation relation (OFDR), that reads

\begin{equation}
R(t,t^\prime)=\beta X(C(t,t^\prime)) \theta(t-t^\prime) \frac{\partial
C(t,t^\prime)}{\partial t^\prime}\ .
\protect\label{OFDR}
\end{equation}
We can relate the previous formula, eq. (\ref{OFDR}), with observable
quantities like the magnetization. The magnetization in the dynamics
is a function of the time and a functional of the magnetic field (that
is itself a function of the time: $h(t)$) and so we can denote it
$m[h](t)$. Using the functional Taylor expansion we can write

\begin{equation}
m[h](t)=m[0](t)+\int_{-\infty}^\infty \d t^\prime ~ \left.\frac{\delta
m[h](t)}{\delta h(t^\prime)}\right|_{h(t)=0} h(t^\prime) + {\rm O}(h^2) \ .
\end{equation}
We define  the response function

\begin{equation}
R(t,t^\prime) \equiv \left. \frac{\delta
m[h](t)}{\delta h(t^\prime)} \right|_{h(t)=0} \ ,
\end{equation}
and using the fact 
that in an Ising spin glass $m[0](t)=0$, we obtain 

\begin{equation}
m[h](t)=\int_{-\infty}^\infty \d t^\prime ~ R(t,t^\prime) 
h(t^\prime) +{\rm O}(h^2)\ .
\end{equation}
Using causality we can reduce the range of the integration to
$(-\infty, t)$:

\begin{equation}
m[h](t)=\int_{-\infty}^t \d t^\prime ~ R(t,t^\prime) 
h(t^\prime) +{\rm O}(h^2)\ .
\end{equation}
This is nothing but that the linear-response theorem if we neglect the terms
proportional to $h^2$.

By applying the OFDR we obtain the dependence of the magnetization
with time in a generic time-dependent magnetic field (with a small
strength), $h(t)$,\footnote{ The symbol $\simeq$ means that the
equation is valid in the region where linear-response holds.}

\begin{equation}
m[h](t)\simeq \beta \int_{-\infty}^t \d t^\prime ~ X[C(t,t^\prime)]  
\frac{\partial C(t,t^\prime)}{\partial t^\prime} 
h(t^\prime) \ .
\end{equation}
Now, we can perform the following experiment. We let the system to
evolve in absence of magnetic field from $t=0$ to $t=t_w$, and then we
turn on a constant magnetic field, $h_0$:
$h(t)=h_0\theta(t-t_w)$.\footnote{ 
  Franz and Rieger~\cite{FRARIE} used a different magnetic
  field function in their study of the fluctuation-dissipation theorem:
  $h_{\rm FR}(t)=h_0 \theta(t_w-t)$.} 
Finally, with our choice of the magnetic field, we can write\footnote{
  We ignore in our notation the fact that $m[h](t)$ depends on $t_w$.}

\begin{equation}
m[h](t)\simeq h_0 \beta \int_{t_w}^t \d t^\prime ~ X[C(t,t^\prime)]  
\frac{\partial C(t,t^\prime)}{\partial t^\prime}\ ,
\protect\label{mag_1}
\end{equation}
and by  performing  the change of variables $u=C(t,t^\prime)$,
equation (\ref{mag_1}) reads

\begin{equation}
m[h](t)\simeq h_0 \beta \int_{C(t,t_w)}^{1} \d u ~ X[u] \ ,
\protect\label{mag_2}
\end{equation}
where we have used the fact that $C(t,t) \equiv 1$ (we work with Ising
spins).
In the equilibrium regime (FDT holds, $X=1$) we must obtain

\begin{equation}
m[h](t)\simeq h_0 \beta \left( 1-C(t,t_w)\right)\ ,
\protect\label{mag_fdt}
\end{equation}
i.e. $m[h](t) T/h_0$ is a linear function of $C(t,t_w)$ with slope
$-1$.

The link with the static is the following. In the limit $t, t_w\to
\infty$ with $C(t,t_w) = q$, $X(C) \to x(q)$, where $x(q)$ is given by

\begin{equation}
x(q)=\int_{0}^q \d q^\prime ~P(q^\prime)\ ,
\protect\label{x_q}
\end{equation}
where $P(q)$ is the equilibrium probability distribution of the
absolute value of the overlap. Obviously $x(q)$ is equal to 1 for all
$q > q_{\rm EA}$, and we recover FDT.

For future convenience, we define 

\begin{equation}
S(C)\equiv \int_C^1 \d q ~x(q) = \int_C^1 \d q ~\int_{0}^q \d
q^\prime ~ P(q^\prime)\ . 
\protect\label{s_c}
\end{equation}
or equivalently

\be
P(C)=-\frac{\d^2 S(C)}{\d^2 C}\ .
\label{pq}
\ee  
In the limit where $X \to x$ we can write eq. (\ref{mag_2}) as

\be
\frac{m[h](t)\,T}{h_0} \simeq S(C(t,t_w)) \ ,
\protect\label{final}
\ee
for large $t_w$. The main goal of this paper is to test this 
last relation (eq. (\ref{final})).

\section{\protect\label{S_FD3}$3D$ Results}

The scheme of our off-equilibrium simulations has been the following.
In a run without magnetic field we compute the autocorrelation
function. We perform a second run where from $t=0$ until $t=t_w$ the
magnetic field is zero and then for $t \ge t_w$ we turn on an uniform
magnetic field of strength $h_0$. The starting configurations were
always chosen at random (i.e. we quench the system suddenly from
$T=\infty$ to the simulation temperature $T$).

We have done a first simulation with $h_0=0.1$ and $t_w=10^5$ , with a
maximum time of $5\cdot 10^6$. A second simulation was done with a
smaller magnetic field, in order to control that linear-response
works: $h_0=0.05$ and $t_w=10^4$ with the same maximum time. The
lattice size in both cases was $64$, the number of samples $4$ and
$T=0.7$ (inside the spin glass phase, the critical temperature is
close to 1.0 \cite{OUR}).

\begin{figure}[htbp]
\begin{center}
\leavevmode
\epsfysize=250pt
\epsffile{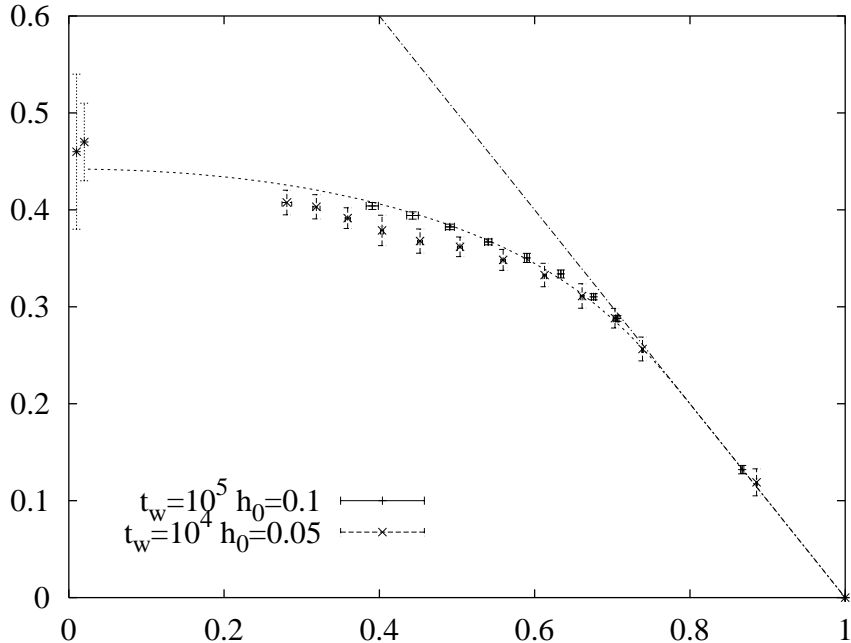}
\end{center}
\caption{$m\ T/h_0$ versus $\!C$ with $L=64$ and $T=0.7$ for the three
dimensional Ising spin glass. The curve is the function $S(C)$
obtained from the equilibrium data. The straight line is the FDT
prediction. We have plotted the data of the two runs: $t_w=10^5$,
$h_0=0.1$  and $t_w=10^4$, $h_0=0.05$.}
\protect\label{fig:gfdt}
\end{figure}

We show in figure (\ref{fig:gfdt}) the numerical results, $m T/h_0$
against $C(t,t_w)$. We have plotted also a straight line with slope
$-1$ in order to control where the FDT is satisfied.

We have also plotted the function $S(C)$, see eq. (\ref{s_c}),
obtained at equilibrium (i.e. using the equilibrium probability
distribution of the overlaps, $P(q)$) by means of a simulation of a
$16^3$ lattice using parallel tempering~\cite{HUKUNEMOTO,ENZO,BOOK}.
We have simulated, with the help of the APE100
supercomputer~\cite{APE}, $900$ samples of a $ L=16$ lattice using the
parallel tempering method simulating $23$ temperatures, from $T=1.8$
down to $T=0.7$ with a step of $0.05$. In order to control the
thermalization we have checked that the $P(q)$ is completely symmetric
in $q$. We have used $10^6$ of sweeps (done of one Metropolis sweep
and one exchange of temperatures) to thermalize and another $10^6$
sweeps (Metropolis+Exchange) to measure (a detail analysis of the
static of the three dimensional Ising spin glass will be presented
elsewhere \cite{OUR}).

Finally we have plotted two points, in the left of the figure, that
are obtained with the infinite time extrapolation of the magnetization
assuming a law

\begin{equation}
m(t)=m_\infty + \frac{A}{t^B} \ ,
\end{equation}
with  $B=0.18(6)$ and $m_\infty T/h_0 =0.46(8)$ in the $h_0=0.05$ run,
and $B=0.21(7)$ and $m_\infty T/h_0 =0.47(4)$ in the $h_0=0.1$
run. The agreement between the two $T m_\infty/h_0$ results is very
good. In the statistical error there are (almost) no differences
between the numerical curves corresponding to the two runs.

From this figure we can estimate the order parameter at this
temperature, that is precisely where the numerical curve and the
straight line with $-1$ slope begin to be different, i.e. where the
violation of FDT starts. We can so estimate $q_{\rm EA}\simeq 0.68$.
We can relate this number with the estimate of $q_{\rm EA}=0.70(2)$
obtained in reference \cite{INMAPARU} using equilibrium simulations.
It is clear that the agreement is very good.

Surprisingly the $S(C)$ curve fit very well the numerical data even in
the region where FDT does not hold, i.e. the equilibrium distribution
determines where begins the violation of the FDT and moreover the
function $x(C)$ is very similar to $X(C)$ even in the very
off-equilibrium regime, in the whole range of $C$. For instance
$S(0)=0.45$ to compare with the off-equilibrium data $T
m_\infty/h_0=0.47(4)$.

In this case we have been able to control down to $C \simeq 0.28$, but
with an optimal combination of $h_0$ and $t_w$ it should be possible
to reach the region of smaller $C$. In any case the infinite time
extrapolation of $m T/h_0$ gives us the final point of the $S(C)$ and so
it should not be difficult to re-construct (by means of educated fits)
the curve $S(C)$ in the region of small $C$.

This analysis implies that the Ansatz $X(t,t^\prime)=X(C(t,t^\prime))$
is correct in finite dimensional spin glasses and that
eq.(\ref{final}) holds in the three dimensional Ising spin glass even
for intermediate waiting times.

\section{\protect\label{S_FD4}$4D$ Results}

In this section we study in detail the scaling properties of the
function $X(T,C)$ and its dependence on the waiting time. We have used
the same procedure as in the three dimensional runs.

\begin{figure}
\begin{center}
\leavevmode
\centering\epsfig{file=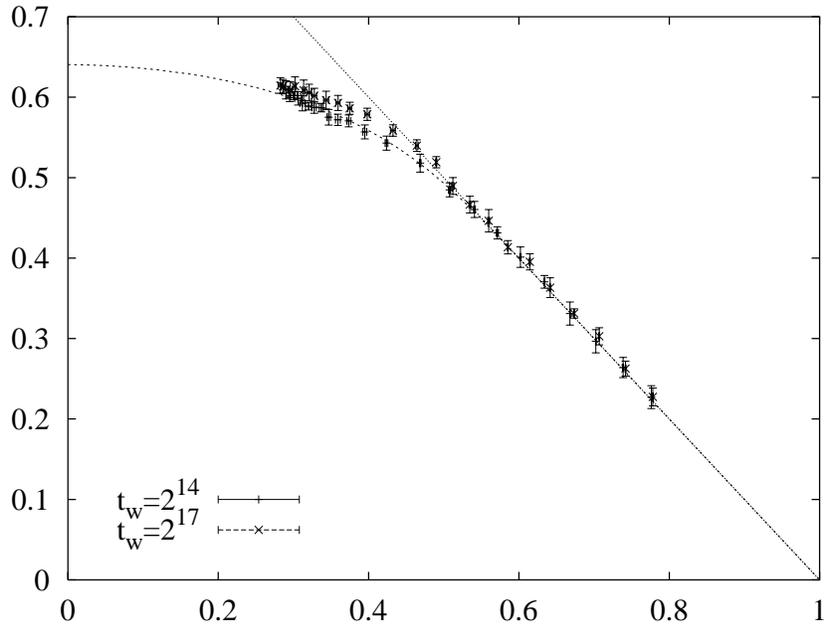,width=0.75\linewidth}
\end{center}
\protect\caption{$m\ T/h_0$ versus $\!C$ with $L=32$ and $T=1.35$
for the four dimensional Ising spin glass. The curve is the function
$S(C)$ obtained from the equilibrium data. The straight line is the
FDT prediction. Here $h_0=0.1$.}
\protect\label{F_stat}
\end{figure}

For the static measurements we have simulated an $L=8$ lattice using
the parallel tempering method. We have simulated 1536 samples in a
range of temperatures that goes from $T=1.35$ to $1.95$ with a step of
$0.05$ (we remark that the transition temperature is 1.80
\cite{4DIM}). We have performed $10^5$ sweeps (Metropolis+Exchange) 
 to thermalize and we have measured, using Metropolis+Exchange,
during $10^5$ sweeps. This takes around one month of the parallel
computer APE100~\cite{APE}. We have checked that thermalization was
achieved by analyzing the symmetry of the overlap probability
distribution. From these simulations we have extracted the function
$S(C)$ shown in figure~\ref{F_stat}.

For the dynamical measurements we have performed off-equilibrium
simulation using the same procedure that we have written in the
previous section.

We take few samples (6 in the present case) of a very large system
($L=24$ and \mbox{$L=32$}) such that it cannot thermalize in any
computer accessible time. We have measured the correlation (runs
without magnetic field) and the response functions of the system for
various waiting times ($t_w = 2^8, 2^{11}, 2^{14}, 2^{17}$) verifying
that for increasing $t_w$ the data of $m T/h_0$ versus $C(t,t_w)$,
plotted in figure~\ref{F_stat}, collapse on a single curve loosing the
dependence on the waiting time. We have simulated almost all the runs
with $h_0=0.1$: only in one  run of a $L=32$ lattice at $T=1.0$ we have
put $h_0=0.05$.

The clear agreement between the static and dynamical data supports
(again) the correctness of the theoretical hypothesis. Nevertheless
the data for the largest waiting time lie a little above the static
curve. We justify this discrepancy noting that in a numerical
simulation of a relatively small volume ($L=8$ in our case) the delta
function for $q=q_{\rm EA}$ in the $P(q)$ is replaced by a quite broad
peak. This effect smoothes the cusp we expect in $S(C)$ at the value
$C=q_{\rm EA}$ lowering the numerical curve with respect to the right
one. In the three dimensional case the data obtained from the
simulation of $16^3$ lattice are very close to asymptotic values (by
comparing, for instance, with $S(C)$ obtained in $8^3$ and $12^3$
lattices \cite{OUR}).

Once we have verified that we can obtain information on the overlap
distribution function $P(q)$ (measuring the linear response of a large
system kept in the out of equilibrium regime) we have performed a
systematic study in the whole frozen phase.

We want to stress that the data from the $L=24$ and the $L=32$ systems
coincide within the errors, suggesting that our results are not
affected by finite size bias. Anyhow, we present data from
both the lattice sizes.

\begin{figure}
\begin{center}
\leavevmode
\centering\epsfig{file=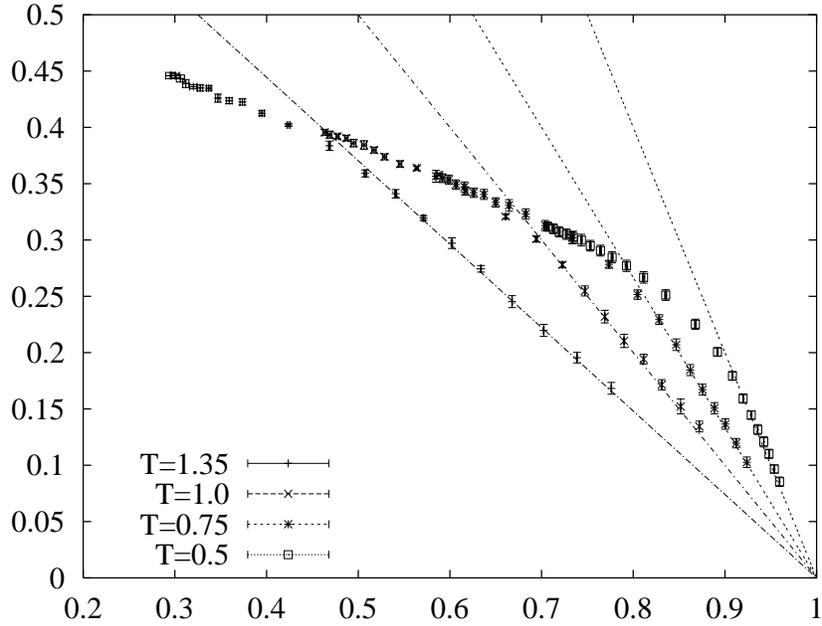,width=0.75\linewidth}
\end{center}
\protect\caption{$m/h_0$ versus $\!C$ with $L=32$ and different
temperatures for the four dimensional Ising spin glass. 
The lines are the FDT regime: $(1-C)/T$. Note how the
data stay on a single curve when they leave the straight line (the FDT
regime). Here $h_0=0.1$.}
\protect\label{F_PAT}
\end{figure}

In figure~\ref{F_PAT} we plot the integrated response against the
correlation function for different temperatures. The straight lines
($m/h_0 = (1-C)/T$) represent the quasi equilibrium regime in which
the system stays while $C>q_{\rm EA}$. Note how the data measured in
the regime where $C<q_{\rm EA}$ collapse on a single curve
independently of the temperature.

We can understand this fact calling a hypothesis that was developed in
the study of the $P(q)$ in the Mean Field approximation by one of the
authors (G.P.) and Toulouse~\cite{PAT,PTV}. 

It assumes that the order
parameter $q(x,T)$~\cite{MEPAVI}, in the Mean Field theory, is a
function of the ratio $x/T$ for $q<q_{\rm EA}$. This imply that we can
write, in this approximation,
\be
x(q,T) = \left\{
\begin{array}{cl}
T \tilde{x}(q) & {\rm for} \;\; q \le q_{\rm EA}(T) \ ,\\
1 & {\rm for} \;\; q > q_{\rm EA}(T) \ ,
\end{array}
\right.
\protect\label{x_qT}
\ee
and, integrating $x(q,T)/T$, we obtain the relation between $m/h_0$
and $C$
\be
\frac{m}{h_0} = \frac{S(C)}{T} = \left\{
\begin{array}{ll}
{\displaystyle\int_C^{q_{\rm EA}}\tilde{x}(q)\d q}+(1-q_{\rm EA})/T &
{\rm for} \;\; C \le q_{\rm EA}(T) \ , \\
(1-C)/T & {\rm for} \;\; C > q_{\rm EA}(T) \ .
\end{array}
\right.
\protect\label{s_cm}
\ee
The terms in the r.h.s. of eq.(\ref{s_cm}) describe the two regimes
present in figure~\ref{F_PAT}: the first gives an expression for the
curve followed by the data in the off-equilibrium regime, while the
second is the straight line (FDT regime).

In the next paragraphs we will show that this
hypothesis~\cite{PAT,PTV} also implies that the off-equilibrium part
is independent of the temperature (i.e. in the region where $C<q_{\rm
EA}$, $[m/h_0](C)$ is independent of the temperature). Using that the
magnetic susceptibility is one in the spin glass phase and with the
help of eq.(\ref{x_qT}) it is possible, with a little algebra, to show
that
\be
\frac{1-q_{\rm EA}(T)}{T} + [1 - T \tilde{x}(q_{\rm EA}(T))]
\frac{\d q_{\rm EA}(T)}{\d T} = 0 \ .
\protect\label{qea}
\ee

Now is very easy to demonstrate that the curves describing the
off-equilibrium regime (\mbox{$C \le q_{\rm EA}(T)$} in
eq.(\ref{s_cm})) do not depend on the temperature. By deriving the
curve expression with respect to $T$ we obtain

\be
\frac{\d}{\d T} \left[\frac{m}{h_0}\right]=\frac{\d}{\d T} 
\left[\frac{S(C)}{T}\right] = \tilde{x}(q_{\rm EA}(T))
\frac{\d q_{\rm EA}(T)}{\d T} - \frac1T \frac{\d q_{\rm EA}(T)}{\d T}
- \frac{1-q_{\rm EA}(T)}{T^2} = 0 \ ,
\ee
where in the last equality we have made use of eq.(\ref{qea}). So
we have verified that the first expression in eq.(\ref{s_cm}) does not
depend on $T$. We finally write that for $C\to q_{\rm EA}^-$ the
hypothesis ~\cite{PAT,PTV} implies
\be
S(C) \simeq \sqrt{1-C} \ .
\label{hypo}
\ee

At this point we have seen that Mean Field predicts qualitatively the
behavior plotted in figure \ref{F_PAT} for a finite dimensional spin
glass. Now we will examine quantitatively the data of figure
\ref{F_PAT}

\begin{figure}
\begin{center}
\leavevmode
\centering\epsfig{file=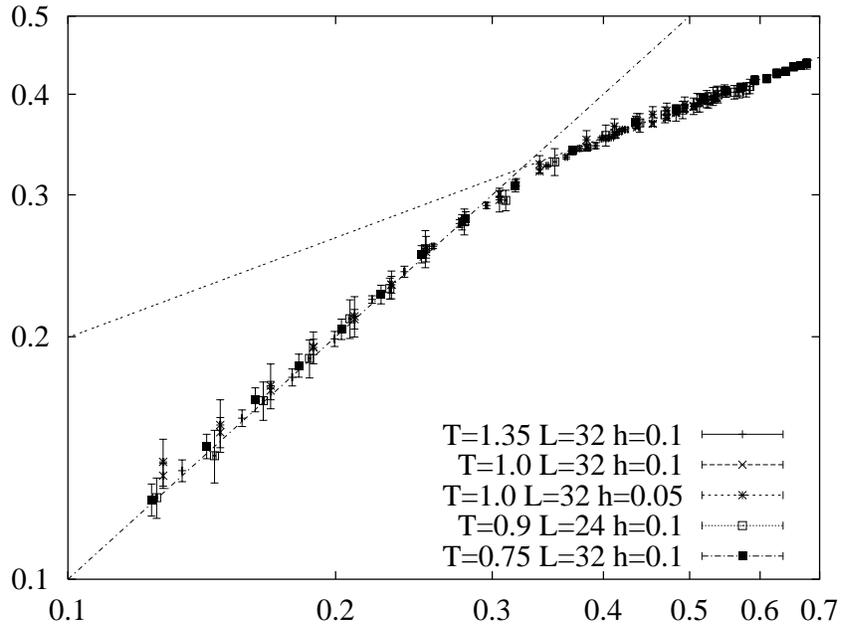,width=0.75\linewidth}
\end{center}
\protect\caption{$ (m T/ h_0)\ T^{-\phi}$ versus $(1-C)\ T^{-\phi}$
with $\phi = 1.7$. Note that in the plot we have included data
measured on different lattices and in presence of different
magnetic fields. In the FDT regime (left part of the figure) the
factor $T^{-\phi}$ has no effect because in this region $m T
/h_0=1-C$. The off-equilibrium regime (right part of the figure)
follows a power law with power $B=0.41$.}
\protect\label{F_SCALED}
\end{figure}

For $C<q_{\rm EA}$ we have seen (figure~\ref{F_PAT}) that the
numerical data can be approximated 
by a power law of the variable
$1-C$
\be
\frac{m T}{h_0} = \left\{
\begin{array}{ll}
T A(1-C)^B & {\rm for} \;\; C \le q_{\rm EA}(T) \ ,\\
1-C & {\rm for} \;\; C > q_{\rm EA}(T) \ ,
\end{array}
\right.
\protect\label{scaling}
\ee
with $A\simeq 0.52$ and $B\simeq 0.41$ (not very far from the Mean
Field behavior, $(1-C)^{1/2}$). Multiplying both sides of the previous
expression by $T^{-1/(1-B)}$ we have
\be
\frac{m T}{h_0} T^{-\frac{1}{1-B}} = \left\{
\begin{array}{rcll}
T^{-\frac{B}{1-B}} A (1-C)^B & = & A \left[(1-C) T^{-\phi}\right]^B &
{\rm for} \;\; C \le q_{\rm EA}(T) \ ,\\
T^{-\frac{1}{1-B}} (1-C) & = & (1-C) T^{-\phi} &
{\rm for} \;\; C > q_{\rm EA}(T) \ ,
\end{array}
\right.
\ee
where we have introduced $\phi=1/(1-B)\simeq 1.7$ for
convenience. Doing so we can rescale the data for all the temperatures
on a single curve like the one shown in figure~\ref{F_SCALED}.

The good scaling of data (figure \ref{F_SCALED}) obtained with
different magnetic fields is a confirmation that we are working in the
linear response regime. It should be noted also the absence of
different finite size effects for the lattices we have considered
($24^4$ and $32^4$).

\section{\protect\label{S_CONCLU}Conclusions}

In this paper we have found that the violation of the
fluctuation-dissipation theorem in finite dimensional spin glasses
follows the lines of the violation of the theorem in Mean Field
models.

We have also found that the function that determine the violation is
given, since intermediate waiting times, by the double integral of the
probability distribution of the overlap calculated at equilibrium.

This fact gives us a further confirmation that the Ansatz used in
references \cite{CUKU} are correct even in finite dimensional models
(i.e. $X$ depends only on $C$, as it was established in reference
\cite{FRARIE}). We have also obtained that the violation of the
theorem is given by the static (i.e. we can express $X(C)$ as a
function of static quantities).

Moreover we have seen that by controlling the scaling of the waiting
times it is possible to construct the $X(C)$ curve without doing
equilibrium simulations. Also these curves provide us an useful and
precise method to compute the Edward-Anderson order parameter.

The form of the $X(C)$ function is very different from that of droplet
(a distinguish ferromagnet), ferromagnetic and one step replica
symmetry breaking systems \cite{GIORGIO_GLASS}, and so we have
obtained another evidence that the finite dimensional Ising spin
glasses cannot be described by the droplet model.

Finally we have studied the scaling properties of $X(C)$ finding that
it is possible parameterize it using static Mean Field analytical
results. It gives us a further evidence of spontaneously broken replica
symmetry (infinite steps of replica symmetry breaking).

\section{\protect\label{S_ACKNOWLEDGES}Acknowledgments}

J.~J.~Ruiz-Lorenzo is supported by an EC HMC (ERBFMBICT950429) grant.


\newpage

\begin{thebibliography}{100}

\bibitem{OUT} J.-P. Bouchaud, L. F. Cugliandolo, J. Kurchan, M. Mezard.
``Out of equilibrium dynamics in spin-glasses and other glassy
systems'' in ``Spin Glasses and Random Fields", edited by
P. Young. Word Scientific (Singapore 1997).  cond-mat/9702070

\bibitem{BOOK} E. Marinari, G. Parisi and
J. J. Ruiz-Lorenzo. ``Numerical Simulations of Spin Glass Systems'' in
``Spin Glasses and Random Fields", edited by P. Young. Word Scientific
(Singapore 1997).  cond-mat/9701016.

\bibitem{MEPAVI} M. Mezard, G. Parisi and M.A. Virasoro, {\em Spin
Glass Theory and Beyond.} (World Scientific, Singapore, 1987).

\bibitem{GIORGIO_GLASS} G. Parisi, Phys. Rev. Lett. (in
press). cond-mat/9703219.

\bibitem{MAPARURI} E. Marinari, G. Parisi, J. J. Ruiz-Lorenzo and
F. Ritort. Phys. Rev. Lett {\bf 76}, 843 (1996).

\bibitem{4DIM} G. Parisi, F. Ricci Tersenghi and J. J. Ruiz-Lorenzo,
J. Phys. A: Math. Gen. {\bf 29}, 7943 (1996).

\bibitem{6DIM} G. Parisi, P. Ranieri, F. Ricci Tersenghi and
J. J. Ruiz-Lorenzo, J. Phys. A: Math. Gen. {\bf 29}, 7115 (1997).

\bibitem{CUKU} L. F. Cugliandolo and J. Kurchan, Phys.  Rev.  Lett.  
{\bf 71}, 173 (1993); Philosophical Magazine {\bf 71}, 501 (1995); J. 
Phys.  A: Math.  Gen.  {\bf 27}, 5749 (1994).

\bibitem{BCKP} A. Baldassarri, L. F. Cugliandolo, J. Kurchan and
G. Parisi, J. Phys. A: Math. Gen. {\bf 28}, 1831 (1995).

\bibitem{FRARIE} S. Franz and H. Rieger, J. of Stat. Phys. {\bf 79},
749 (1995).

\bibitem{OUR} E. Marinari, G. Parisi and J. J. Ruiz-Lorenzo. In
preparation.

\bibitem{HUKUNEMOTO} K. Hukushima and K. Nemoto, J. Phys. Soc. Japan,
65, No 6, 1604 (1996).

\bibitem{ENZO}E. Marinari, {\em Optimized Monte Carlo Methods},
cond-mat/ 9612010.

\bibitem{APE} C. Battista {\em et al}. Int. J. High Speed Comp. {\bf
5}, 637 (1993).

\bibitem{INMAPARU} D. I\~niguez, E. Marinari, G. Parisi and
J. J. Ruiz-Lorenzo, Journal of Physics A (in press). cond-mat/9707050.

\bibitem{PAT} G. Parisi and G. Toulouse, J. de Phys. Lett. {\bf 41}, L361
(1980).

\bibitem{PTV} G. Parisi, G. Toulouse and Vannimenus, J. de Phys. {\bf
42} 565 (1981).

\end{thebibliography}
\end{document}